%% file: phasetransition_short_final_clean.tex
\documentclass[12pt]{article}

 \usepackage[utf8]{inputenc}
 \usepackage{amsmath}
 \usepackage{amssymb}
 \usepackage{amsthm}
 \usepackage{nicefrac}
 \usepackage{vmargin} 
 \usepackage{graphicx}
 \usepackage{color}
 \usepackage{epsfig}
 \usepackage[only,llbracket,rrbracket]{stmaryrd}
 \usepackage{hyperref}
 \usepackage{cite}

\newtheorem{definition}{Definition}[section]
\newtheorem{hypothesis}{Hypothesis}[section]
\newtheorem{theorem}{Theorem}
\newtheorem{proposition}{Proposition}[section]
\newtheorem{lemma}{Lemma}

\renewcommand{\d}{\mathrm d}  

\title{A note on phase transitions for the Smoluchowski equation with dipolar potential}

\author{Pierre Degond$^{(1,2)}$, Amic Frouvelle$^{(3)}$,  Jian-Guo Liu$^{(4)}$}
\date{}
\begin{document}
\maketitle

\vspace{0.2 cm}

\begin{center}
1-Université de Toulouse; UPS, INSA, UT1, UTM ;\\ 
Institut de Mathématiques de Toulouse ; \\
F-31062 Toulouse, France. \\
2-CNRS; Institut de Mathématiques de Toulouse UMR 5219 ;\\ 
F-31062 Toulouse, France.\\
email: pierre.degond@math.univ-toulouse.fr, amic.frouvelle@math.univ-toulouse.fr
\end{center}

\begin{center}
3- CEREMADE-UMR CNRS 7534\\
Universit\'e de PARIS-DAUPHINE\\
75775 PARIS CEDEX 16, France\\
email: frouvelle@ceremade.dauphine.fr
\end{center}

\begin{center}
4- Department of Physics and Department of Mathematics\\
Duke University\\
Durham, NC 27708, USA\\
email: jliu@phy.duke.edu
\end{center}

\begin{abstract}
In this note, we study the phase transitions arising in a modified Smoluchowski equation on the sphere with dipolar potential. This equation models the competition between alignment and diffusion, and the modification consists in taking the strength of alignment and the intensity of the diffusion as functions of the order parameter.

We characterize the stable and unstable equilibrium states. For stable equilibria, we provide the exponential rate of convergence. We detail special cases, giving rise to second order and first order phase transitions, respectively. We study the hysteresis diagram, and provide numerical illustrations of this phenomena.
\end{abstract}

\section{Introduction}
\label{sec-intro}

In this short note, we study the following modified Smoluchowski equation (also called Fokker--Planck equation), for an orientation distribution~$f(\omega,t)$ defined for a time~$t\geqslant0$ and a direction~$\omega\in\mathbb{S}$ (the unit sphere~$\mathbb{S}$ of~$\mathbb{R}^n$) as follows:
\begin{gather}
\label{FP-eq}
\partial_tf = - \nu(|J_f|)\nabla_\omega \cdot (f \, \nabla_\omega(\omega \cdot \Omega_{f}) ) + \tau(|J_f|)\Delta_\omega f =:Q(f),\\
\Omega_f=\frac{J_f}{|J_f|}, \quad J_f(t)=\int_{ \upsilon \in \mathbb{S} }  \upsilon \, f(\upsilon,t) \, \d\upsilon.
\end{gather}
where~$\Delta_\omega$,~$\nabla_\omega\cdot$, and~$\nabla_\omega$ are the Laplace–Beltrami, divergence, and gradient operators on the sphere. The vector~$J_f\in\mathbb{R}^n$ is the first moment associated to~$f$ (the measure on the sphere is the uniform measure such that~$\int_\mathbb{S} \d\upsilon = 1$), and~$\Omega_f\in\mathbb{S}$ represents the mean direction of the distribution~$f$.

The first term of the right-hand side of~\eqref{FP-eq} is an alignment term towards~$\Omega_f$, and the function~$\nu$ represents the strength of this alignment. The function~$\tau$ is the intensity of the diffusion on the sphere.

When~$\tau$ is a constant and~$\nu(|J_f|)=|J_f|$, we can recover the standard Smoluchowski equation on the sphere, with dipolar potential~\cite{frouvelle2012dynamics}. Indeed, the dipolar potential is given by~$K(\omega,\bar{\omega})=-\omega\cdot\bar{\omega}$, and the equation can be recast as:
\begin{equation}
\label{FP-dipolar}
\partial_tf = \nabla_\omega \cdot \Big(f \, \nabla_\omega\big(\int_\mathbb{S}K(\omega\cdot\bar{\omega})f(\bar{\omega},t)\d \bar{\omega}\big) \Big) + \tau\Delta_\omega f.
\end{equation}

Other classical kernels~\cite{maier1958eine,onsager1949effects,doi1999theory} for the study of semi-dilute and concentrated suspensions of polymers are the so-called Maier–Saupe potential~$K(\omega,\bar{\omega})=-(\omega\cdot\bar{\omega})^2$, or the original Onsager potential~$K(\omega,\bar{\omega})=|\omega\times\bar{\omega}|$. In these cases, there are a lot of studies regarding the phase transition phenomenon for equilibrium states~\cite{constantin2004asymptotic,zhang2007stable,fatkullin2005critical,fatkullin2005note,chen2010stationary,liu2005axial,lucia2010exact,wang2008multiple,wang2008unified,zhou2007characterization,zhou2005new}, and in particular a hysteresis phenomenon occurs for the Maier--Saupe potential in dimension~$3$.
The case of the dipolar potential has also been studied precisely~\cite{frouvelle2012dynamics}, with an analysis of the rates of convergence to the equilibrium as time goes to infinity. In this case, there exists a so-called continuous phase transition for a critical threshold~$\tau_c$: when~$\tau<\tau_c$ the solution converges exponentially fast to a non-isotropic equilibrium; when~$\tau>\tau_c$ it converges exponentially fast to the uniform distribution. At the critical value~$\tau=\tau_c$, the solution converges to the uniform distribution at a rate~$t^{-1/2}$.

Here we study the modifications arising when~$\nu$ and~$\tau$ depend on~$|J_f|$, which can be motivated by some biological modeling~(see \cite{degond2008continuum} and references therein). In that case, we cannot use~$\tau$ as a bifurcation parameter anymore. Instead, we will use the initial mass~$\rho$ (a conserved quantity) as the key parameter to study the phase transition. We will assume that:
\begin{hypothesis}~
\label{hyp-regularity}
\begin{enumerate}
\item[(i)] The functions~$\nu$ and~$\tau$ are~$C^1$, with~$\nu(0)=0$ and~$\tau>0$.
\item[(ii)] The function $|J|\mapsto h(|J|)=\frac{\nu(|J|)}{\tau(|J|)}$ is an increasing function. We denote by~$\sigma$ its inverse, i.e.
\[\kappa=h(|J|)\Leftrightarrow|J|=\sigma(\kappa).\]
\end{enumerate}
\end{hypothesis}

The first part of this hypothesis implies that we do not have any singularity of~$Q$ in~\eqref{FP-eq} as~\mbox{$|J_f|\to0$}: if~$|J_f|=0$, we simply have~$Q(f)=\tau(0)\Delta_\omega f$. The second part is made for the sake of simplicity. It leaves enough flexibility to to reveal key behaviors in terms of phase transitions. It would be easy to remove it at the price of an increased technicality. Additionally, it means that when~$f$ is more concentrated in the direction of~$\Omega_f$, the relative strength of the alignment force compared to diffusion is increased as well. This can be biologically motivated by the existence of some social reinforcement mechanism. 

We will see that we can observe a wealth of phenomena, including hysteresis. The purpose of this note is to summarize the analytical results, as well as some numerical simulations which illustrate this phenomena. All the proofs are detailed in~\cite{degond2013phase}, where~\eqref{FP-eq} arises as the spatially homogeneous version of a space-dependent kinetic equation, obtained as the mean-field limit of a self-propelled particle system interacting through alignment. This spatially homogeneous study is crucial to determine the macroscopic behavior of this space-dependent kinetic equation.

\section{General study}
\subsection{Existence and uniqueness}
We first state results about existence, uniqueness, positivity and regularity of the solutions of~\eqref{FP-eq}. Under hypothesis~\ref{hyp-regularity}, we have the following
\begin{theorem}
\label{theorem-existence-uniqueness}
Given an initial finite non-negative measure~$f_0$ in~$H^s(\mathbb{S})$, there exists a unique weak solution~$f$ of~\eqref{FP-eq} such that~$f(0)=f_0$. 
This solution is global in time. 

Moreover,~$f\in C^\infty((0,+\infty)\times\mathbb{S})$, with~$f(t,\omega)>0$ for all positive~$t$, and we have the following instantaneous regularity and uniform boundedness estimates (for~$m\in\mathbb{N}$, the constant~$C$ being independent of~$f$), for all~$t>0$:
\begin{equation*}
\|f(t)\|^2_{H^{s+m}}\leqslant C\left(1+\frac1{t^m}\right)\|f_0\|^2_{H^{s}}.
\end{equation*}
\end{theorem}

For later usage, we define~$\Phi(|J|)$ as an anti-derivative of~$h$:~$\frac{\d \Phi}{\d |J|}=h(|J|)$.
In this case, the dynamics of~\eqref{FP-eq} corresponds to the gradient flow of the following free energy functional:
\[
\mathcal F(f)=\int_\mathbb{S}f\ln f\, \d \omega - \Phi(|J_f|).
\]
Indeed, if we define the dissipation term~$\mathcal D(f)$ by
\[
\mathcal D(f)=\tau(|J_f|)\int_\mathbb{S}f\,|\nabla_\omega(\ln f-h(|J_f|)\,\omega\cdot\Omega_f)|^2\, \d \omega,
\]
we get the following conservation relation:
\begin{equation}
\label{dissipation-free-energy}
\frac{\d}{\d t} \mathcal F(f)=-\mathcal D(f)\leqslant0.
\end{equation}

\subsection{Equilibria}
\label{subsec-equilibria}

We now define the von Mises distribution which provides the general shape of the non-isotropic equilibria of~$Q$.

\begin{definition}
\label{def-vonMises}
The von Mises distribution of orientation~$\Omega\in\mathbb{S}$ and concentration parameter~$\kappa\geqslant0$ is given by:
\begin{equation}
\label{eq-vonMises}
M_{\kappa\Omega}(\omega) =\frac{e^{\kappa\, \omega \cdot \Omega}}{\int_\mathbb{S} e^{\kappa\, \upsilon \cdot \Omega}\, \d\upsilon}.
\end{equation}
The order parameter~$c(\kappa)$ is defined by the relation
\[
J_{M_{\kappa\Omega}}=c(\kappa)\Omega,
\]
and has expression:
\begin{equation}
\label{eq-c}
c(\kappa)=\frac{\int_0^\pi \cos\theta \, e^{\kappa \cos\theta}\sin^{n-2}\theta \, \d\theta}{\int_0^\pi e^{\kappa \cos\theta}\sin^{n-2}\theta \, \d\theta}.
\end{equation}
\end{definition}
\noindent The concentration parameter~$c(\kappa)$ defines a one-to-one correspondence $\kappa\in[0,\infty)\mapsto c(\kappa)\in[0,1)$. The case~$\kappa=c(\kappa)=0$ corresponds to the uniform distribution, while when~$\kappa$ is large (or~$c(\kappa)$ is close to~$1$), the von Mises distribution is closed to a Dirac delta mass at the point~$\Omega$.

Some comments are necessary about the interval of definition of~$\sigma$. First note that, under hypothesis~\ref{hyp-regularity},~$h$ is defined from~$[0,+\infty)$, with values in an interval~$[0,\kappa_{max})$, where we may have~$\kappa_{max}=+\infty$. So~$\sigma$ is an increasing function from~$[0,\kappa_{max})$ onto~$\mathbb{R}_+$.
Moreover, for later usage, we can define
\begin{equation}
\label{def-rho-c}
 \tau_0=\tau(0)>0,\quad \text{ and } \quad \rho_c=\lim_{|J|\to0}\frac{n|J|}{h(|J|)}=\frac{n\tau_0}{\nu'(0)} 
\end{equation}
where~$\rho_c>0$ may be equal to~$+\infty$, and where we recall that~$n$ denotes the dimension.

The equilibria are given by the following proposition:

\begin{proposition}
\label{prop-equilibria}
The following statements are equivalent:
\begin{enumerate}
\item[(i)] $f\in C^2(\mathbb{S})$ and~$Q(f)=0$.
\item[(ii)] $f\in C^1(\mathbb{S})$ and~$\mathcal D(f)=0$.
\item[(iii)] There exists~$\rho\geqslant0$ and~$\Omega\in\mathbb{S}$ such that~$f=\rho M_{\kappa\Omega}$, where~$\kappa\geqslant0$ satisfies the compatibility equation:
\begin{equation}
\label{eq-compatibility}
\sigma(\kappa)=\rho c(\kappa).
\end{equation}
\end{enumerate}
\end{proposition}

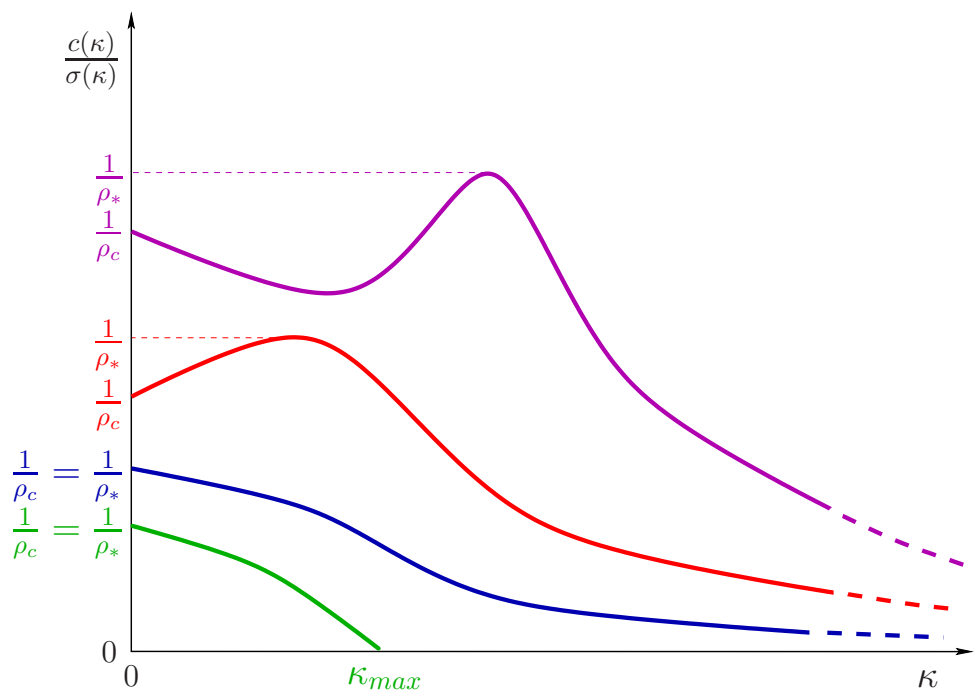
\begin{figure}[!h]
\begin{center}
\input{shapes_c_sigma_versus_kappa.pspdftex}
\caption{The green, blue, red and purple curves correspond to various possible profiles for the function~$\kappa\mapsto\frac{c(\kappa)}{\sigma(\kappa)}$.}
\label{fig-shapes}
\end{center}
\end{figure}

Let us first remark that the uniform distribution, corresponding to~$\kappa=0$ is always an equilibrium. Indeed, we have~$c(0)=\sigma(0)=0$ and~\eqref{eq-compatibility} is satisfied. However, Proposition~\ref{prop-equilibria} does not provide any information about the number of the non-isotropic equilibria. Indeed, equation~\eqref{eq-compatibility} can be recast into:
\begin{equation}
\label{eq-compat2}
\frac{c(\kappa)}{\sigma(\kappa)}=\frac1{\rho},
\end{equation}
which is valid as long as~$\sigma\neq0$. We know that~$\sigma$ is an increasing unbounded function from its interval of definition~$[0,\kappa_{max})$ onto~$[0,+\infty)$, and thanks to hypothesis~\ref{hyp-regularity} and to~\eqref{def-rho-c}, we know that~$\sigma(\kappa)\sim \frac{\rho_c}{n} \kappa$ as~$\kappa\to0$ (if~$\rho_c<+\infty$). So since~$c(\kappa)\sim \frac1{n}\kappa$ as~$\kappa\to0$ (see for instance~\cite{frouvelle2012dynamics}), we have the two following results (also valid in the case~$\rho_c=+\infty$):
\begin{equation}
\frac{c(\kappa)}{\sigma(\kappa)}\to\frac1{\rho_c}\text{ as }{\kappa\to0} \quad \text{ and } \quad \frac{c(\kappa)}{\sigma(\kappa)}\to0\text{ as }{\kappa\to\kappa_{max}}.
\end{equation}
We deduce that this function reaches its maximum, and we define
\begin{equation}
\label{def-rho-star}
\rho_*=\min_{\kappa\in\mathbb{R}_+}\frac{\sigma(\kappa)}{c(\kappa)}.
\end{equation}
For~$\rho<\rho_*$, the only solution to the compatibility condition is~$\kappa=0$, and the only equilibrium is the uniform distribution~$f=\rho$.
Except from these facts, we have no further direct information of this function~$\kappa\mapsto c(\kappa)/\sigma(\kappa)$, since~$c$ and~$\sigma$ are both increasing.
Figure~\ref{fig-shapes} depicts some examples of the possible shapes of the function~$\kappa\mapsto c(\kappa)/\sigma(\kappa)$.

We see that depending on the value of~$\rho$, the number of families of non-isotropic equilibria, given by the number of positive solutions of the equation~\eqref{eq-compat2}, can be zero, one, two or even more.
We now turn to the study of the stability of these equilibria, through the study of the rates of convergence. 

\subsection{Rates of convergence to equilibrium}

The main tool to prove convergence of the solution to a steady state is LaSalle’s principle, that we recall here (the proof follows exactly the lines of~\cite{frouvelle2012dynamics}).
By the conservation relation~\eqref{dissipation-free-energy}, we know that the free energy~$\mathcal F$ is decreasing in time (and bounded from below since~$|J|$ is bounded). LaSalle’s principle states that the limiting value of~$\mathcal F$ corresponds to an~$\omega$-limit set of equilibria:

\begin{proposition}
\label{prop-lasalle}LaSalle’s invariance principle.

Let~$f_0$ be a positive measure on the sphere~$\mathbb{S}$. 
We denote by~$\mathcal F_\infty$ the limit of~$\mathcal F(f(t))$ as~$t\to\infty$, where~$f$ is the solution to the modified Smoluchowski equation~\eqref{FP-eq} with initial condition~$f_0$.

Then the set~$\mathcal E_\infty=\{f \in C^\infty(\mathbb{S}) \text{ s.t. } \mathcal D(f)=0 \text{ and } \mathcal F(f)=\mathcal F_\infty \}$ is not empty.

Furthermore~$f(t)$ converges in any~$H^s$ norm to this set of equilibria (in the following sense):
\begin{equation*}
\lim_{t\to\infty}\, \inf_{g\in\mathcal E_\infty}\|f(t)-g\|_{H^s}=0.
\end{equation*}
\end{proposition}

Since we know the types of equilibria, we can refine this principle to adapt it to our problem:

\begin{proposition}
\label{prop-lasalle-refined}

Let~$f_0$ be a positive measure on the sphere~$\mathbb{S}$, with initial mass~$\rho$.

If no open interval is included in the set~$\{\kappa,\rho c(\kappa)=\sigma(\kappa)\}$, then there exists a solution~$\kappa_\infty$ to the compatibility solution~\eqref{eq-compatibility} such that we have:
\begin{gather*}
\lim_{t\to\infty} |J_f(t)|=\rho c(\kappa_\infty)\\
\forall s\in\mathbb{R}, \lim_{t\to\infty}\,\|f(t)-\rho M_{\kappa_\infty\Omega_f(t)}\|_{H^s}=0.
\end{gather*}
\end{proposition}

This last proposition helps us to characterize the~$\omega$-limit set by studying the single compatibility equation~\eqref{eq-compatibility}.

When~$\kappa=0$ is the unique solution, then this gives us that~$f$ converges to the uniform distribution.
Otherwise, two cases are possible, either~$\kappa_\infty=0$, and~$f$ converges to the uniform distribution, or~$\kappa_\infty\neq0$, and the only unknown behavior is the one of~$\Omega_{f(t)}$. If we are able to prove that it converges to~$\Omega_\infty\in\mathbb{S}$, then~$f$ converges to a fixed non-isotropic steady-state~$\rho M_{\kappa_\infty\Omega_\infty}$.

However, Proposition~\ref{prop-lasalle-refined} does not give information about quantitative rates of convergence of~$|J_f|$ to~$\rho c(\kappa_\infty)$, and of~$\|f(t)-\rho M_{\kappa_\infty\Omega_f(t)}\|_{H^s}$ to~$0$, as~$t\to\infty$. So we now turn to the study of the behavior of the difference between the solution~$f$ and a target equilibrium~$\rho M_{\kappa_\infty\Omega_f(t)}$. 

This study consists in two types of expansion. If we expand the solution around the uniform equilibrium, some simple energy estimates give us exponential convergence when~$\rho<\rho_c$. But when we expand the solution around a non-isotropic equilibrium~$\rho M_{\kappa_\infty\Omega_f(t)}$, we see that the condition of stability is related to the monotonicity of the function~$\kappa\mapsto c(\kappa)/\sigma(\kappa)$. Hence we can see directly on the graph of this function (see examples on Figure~\ref{fig-shapes}) both the number of family of equilibria and their stability: if the function is decreasing, the family is stable. By contrast it is unstable when the function is increasing. When the difference between~$f$ and~$\rho M_{\kappa_\infty\Omega_f(t)}$ converges exponentially fast to~$0$ (on the stable branch), we are able to control the displacement of~$\Omega_f(t)$, which gives convergence to~$\Omega_\infty\in\mathbb{S}$. We then have convergence of~$f$ to a given equilibrium$\rho M_{\kappa_\infty\Omega_\infty}$.

All these results are summarized in the following two theorems. 
In what follows, we say that a constant is a universal constant when it does not depend on the initial condition~$f_0$ (that is to say, it depends only on~$\rho$,~$n$ and the coefficients of the equation~$\nu$ and~$\tau$, and on the exponent~$s$ of the Sobolev space~$H^s$ in which the result is stated). 

\begin{theorem} We have the following instability and exponential stability results around the uniform equilibrium:
\label{thm-uniform}
\begin{itemize}
\item Suppose that~$\rho<\rho_c$. We define
\begin{equation}
\lambda=(n-1)\tau_0(1-\frac{\rho}{\rho_c})>0.
\end{equation}
There exists a universal constant~$C$, such that if~$\|f_0-\rho\|_{H^s}<\frac{\lambda}{C}$, then for all~$t\geqslant0$, we have 
\[
\|f(t)-\rho\|_{H^s}\leqslant\frac{\|f_0-\rho\|_{H^s}}{1-\frac{C}{\lambda}\|f_0-\rho\|_{H^s}}e^{-\lambda t}.
\]
\item If~$\rho>\rho_c$, and if~$J_{f_0}\neq0$, then we cannot have~$\kappa_\infty=0$ in Proposition~\ref{prop-lasalle-refined}: the solution cannot converge to the uniform equilibrium.
\end{itemize}
\end{theorem}

To study the stability around a non-isotropic equilibrium, we fix~$\rho$, and we denote by~$\kappa$ a positive solution to the compatibility equation (we will not write the dependence of~$c$ and~$\sigma$ on~$\kappa$ when there is no possible confusion). We denote by~$\mathcal F_\kappa$ the value of~$\mathcal F(\rho M_{\kappa\Omega})$ (independent of~$\Omega\in\mathbb{S}$).
\begin{theorem}
\label{thm-nonisotropic} 
We have the following instability and exponential stability results when starting close to a non-isotropic equilibrium:
\begin{itemize}
\item Suppose~$(\frac{\sigma}{c})'(\kappa)>0$.
For all~$s>\frac{n-1}2$, there exist universal constants~$\delta>0$ and~$C>0$, such that for any initial condition~$f_0$ satisfying~$\|f_0-\rho M_{\kappa\Omega}\|_{H^s}<\delta$ for some~$\Omega\in\mathbb{S}$, there exists~$\Omega_\infty\in\mathbb{S}$ such that
\[
\|f-\rho M_{\kappa\Omega_\infty}\|_{H^s}\leqslant C\|f_0-\rho M_{\kappa\Omega_{f_0}}\|_{H^s}e^{-\lambda t},
\]
where the rate is given by
\begin{equation}
\label{def-lambda}
\lambda=\frac{c\tau(\sigma)}{\sigma'}\Lambda_\kappa(\frac{\sigma}{c})'.
\end{equation}
The constant~$\Lambda_\kappa$ is the best constant for the following weighted Poincaré inequality (see the appendix of~\cite{degond2012macroscopic} for more details on this constant, which does not depend on~$\Omega$):
\begin{equation}
\label{poincare-lambda}
\int_\mathbb{S}|\nabla_\omega g|^2\,M_{\kappa\Omega}\,\d \omega\geqslant\Lambda_\kappa\Big[\int_\mathbb{S}g^2M_{\kappa\Omega}\,\d \omega - \big(\int_\mathbb{S}g\,M_{\kappa\Omega}\,\d \omega\big)^2\Big].
\end{equation}
\item Suppose~$(\frac{\sigma}{c})'(\kappa)<0$. Then any equilibrium of the form~$\rho M_{\kappa\Omega}$ is unstable, in the following sense:
in any neighborhood of~$\rho M_{\kappa\Omega}$, there exists an initial condition~$f_0$ such that~$\mathcal F(f_0)<\mathcal F_\kappa$. Consequently, in that case, we cannot have~$\kappa_\infty=\kappa$ in Proposition~\ref{prop-lasalle-refined}.
\end{itemize}
\end{theorem}

\section{Second order phase transition}

Let us now focus on the case where we always have~$(\frac{\sigma}{c})'>0$ for all~$\kappa>0$ (see for example the lowest two curves of Figure~\ref{fig-shapes}). In this case, the compatibility equation~\eqref{eq-compat2} has a unique positive solution for~$\rho>\rho_c$. With the results of the previous subsection about stability and rates of convergence, we obtain the behavior of the solution for any initial condition~$f_0$ with initial mass~$\rho$.
\begin{itemize}
\item If~$\rho<\rho_c$, then the solution converges exponentially fast towards the uniform distribution~$f_\infty=\rho$.
\item If~$\rho=\rho_c$, the solution converges to the uniform distribution.
\item If~$\rho>\rho_c$ and~$J_{f_0}\neq0$, then there ~$\Omega_\infty$ such that~$f$ converges exponentially fast to the von Mises distribution~$f_\infty=\rho M_{\kappa\Omega_\infty}$, where~$\kappa>0$ is the unique positive solution to the equation~$\rho c(\kappa)=\sigma(\kappa)$.
\end{itemize}
The special case where~$J_{f_0}=0$ leads to the heat equation~$\partial_tf=\tau_0\Delta_\omega f$. Its solution converges exponentially fast to the uniform distribution, but this solution is not stable under small perturbation of the initial condition. Let us remark that for some particular choice of the coefficients, as in~\cite{frouvelle2012dynamics}, it is also possible to get an algebraic rate of convergence in the second case~$\rho=\rho_c$. For example when~$\sigma(\kappa)=\kappa$, we have~$\|f-\rho\|\leqslant Ct^{-\frac12}$ for~$t$ sufficiently large.

So we can describe the phase transition phenomena by studying the order parameter of the asymptotic equilibrium~$c=\frac{|J_{f_\infty}|}{\rho}$, as a function of the initial density~$\rho$.

We have~$c(\rho)=0$ if~$\rho\leqslant\rho_c$, and~$c$ is a positive continuous increasing function for~$\rho>\rho_c$. In the common situation where~$\frac{c}{\sigma}=\frac{1}{\rho_c}-a\kappa^{\frac1{\beta}} + o(\kappa^{\frac1{\beta}})$ when~$\kappa\to0$, it is easy to see, since~$c(\kappa)\sim\frac1n \kappa$ when~$\kappa\to0$, that we have
\begin{equation}
\label{critical-exponent}
c(\rho)\sim\widetilde{a}(\rho-\rho_c)^\beta,\text{ as }\rho\overset{>}{\to}\rho_c.
\end{equation}
Since~$\frac{c}{\sigma}$ is Lipschitz, we always have~$\beta\leqslant1$. So the first derivative of~$c$ is discontinuous at~\mbox{$\rho=\rho_c$}. This is the case of a second order phase transition (also called continuous phase transition). The critical exponent~$\beta$ can take arbitrary values in~$(0,1]$, as can be seen by taking~$h(|J|)$ such that~$\sigma(\kappa)=c(\kappa)(1+\kappa^{\frac1{\beta}})$.

In general, we have the following practical criterion, which ensures a second order phase transition.
\begin{lemma}
\label{lemma-2nd-order}
If~$\frac{h(|J|)}{|J|}$ is a non-increasing function of~$|J|$, then we have~$(\frac{\sigma}{c})'>0$ for all~$\kappa>0$. In this case, the critical exponent~$\beta$ in~\eqref{critical-exponent}, if it exists, can only take values in~$[\frac12,1]$. \end{lemma}

\section{Hysteresis}
\subsection{Typical example}
We now turn to a specific example, where all the features presented in the stability study can be seen. We focus on the case where~$\nu(|J|)=|J|$, as in~\cite{frouvelle2012dynamics}, but we now take~$\tau(|J|)=1/(1+|J|)$. From the modeling point of view, this occurs in the Vicsek model with vectorial noise (also called extrinsic noise)~\cite{aldana2009emergence,chate2008collective}.

In this case, we have~$h(|J|)=|J|+|J|^2$, so the assumptions of Lemma~\ref{lemma-2nd-order} are not fulfilled, and the function~$\sigma$ is given by~$\sigma(\kappa)=\frac12(\sqrt{1+4\kappa}-1)$.

Expanding~$\frac{c}{\sigma}$ when~$\kappa$ is large or~$\kappa$ is close to~$0$, we get
\[
\frac{c}{\sigma}=\begin{cases}
\frac1n+\frac1n\kappa+O(\kappa^2)&\text{ as }\kappa\to0,\\
\frac1{\sqrt\kappa}+O(\kappa^{-1})&\text{ as }\kappa\to\infty.
\end{cases}
\]
Consequently, there exist more than one family of non-isotropic equilibria when~$\rho$ is close to~$\rho_c=n$ (and $\rho>\rho_c$). 

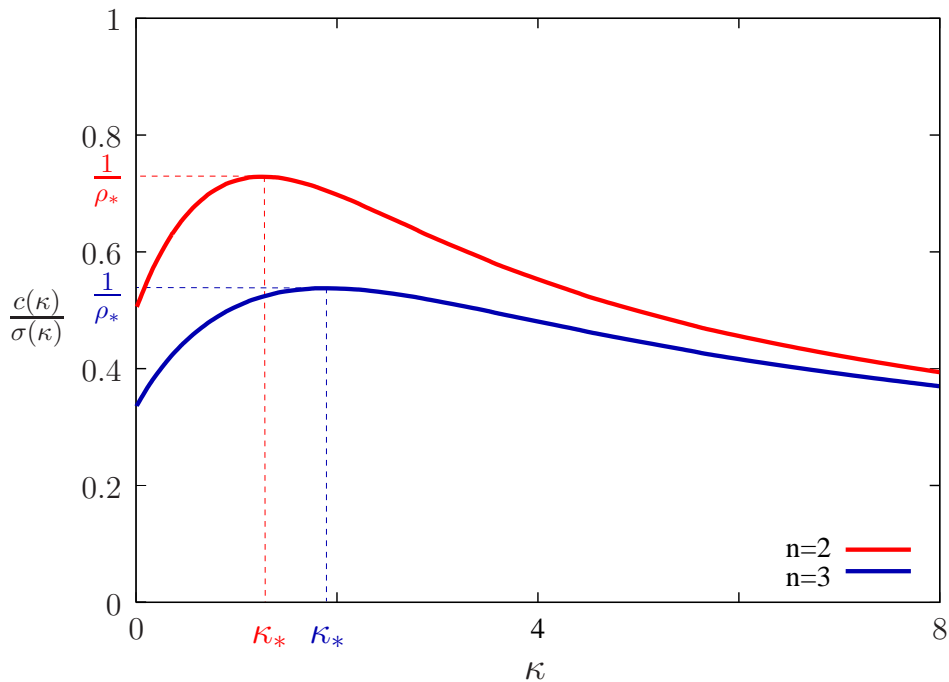
\begin{figure}[!h]
\begin{center}
\input{c_sigma_versus_kappa.pspdftex}
\caption{The function~$\kappa\mapsto\frac{c(\kappa)}{\sigma(\kappa)}$, in dimensions~$2$ and~$3$.}
\label{fig-2d-3d-hysteresis}
\end{center}
\end{figure}

The function~$\kappa\mapsto\frac{c(\kappa)}{\sigma(\kappa)}$ can be computed numerically. The results are displayed in Figure~\ref{fig-2d-3d-hysteresis} in dimensions~$n=2$ and~$n=3$.

We observe the following features:
\begin{itemize}
\item There exists a unique critical point~$\kappa_*$ for the function~$\frac{c}{\sigma}$, corresponding to its global maximum~$\frac1{\rho_*}$ (in dimension~$2$, we obtain numerically~$\rho_*\approx 1.3726$ and~$\kappa_*\approx 1.2619$, in dimension~$3$ we get~$\rho_*\approx 1.8602$ and~$\kappa_*\approx 1.9014$).
\item The function~$\frac{c}{\sigma}$ is strictly increasing in~$[0,\kappa_*)$ and strictly decreasing on~$(\kappa_*,\infty)$.
\end{itemize}

From these properties, it follows that the solution associated to an initial condition~$f_0$ with mass~$\rho$ can exhibit different types of behavior, depending on the three following regimes for~$\rho$.
\begin{itemize}
\item If~$\rho<\rho_*$, the solution converges exponentially fast to the uniform equilibrium~$f_\infty=\rho$.
\item If~$\rho_*<\rho<n$, there are two families of stable solutions: either the uniform equilibrium~$f=\rho$ or the von Mises distributions of the form~$\rho M_{\kappa\Omega}$, for~$\Omega\in\mathbb{S}$ where~$\kappa$ is the unique solution with~$\kappa>\kappa_*$ of the compatibility equation~\eqref{eq-compatibility}. 
If~$f_0$ is sufficiently close to one of these equilibria, there is exponential convergence to an equilibrium of the same family.

The von Mises distributions of the other family (corresponding to solution of~\eqref{eq-compatibility} such that~$0<\kappa<\kappa_*$) are unstable in the sense given in Theorem~\ref{thm-nonisotropic}.
\item If~$\rho>n$ and~$J_{f_0}\neq0$, then there exists~$\Omega_\infty\in\mathbb{S}$ such that~$f$ converges exponentially fast to the von Mises distribution~$\rho M_{\kappa\Omega_\infty}$, where~$\kappa$ is the unique positive solution to the compatibility equation~$\rho c(\kappa)=\sigma(\kappa)$. 
\end{itemize}
At the critical point~$\rho=\rho_*$, the uniform equilibrium is stable (and for any initial condition sufficiently close to it, the solution converges exponentially fast to it), but the stability of the family of von Mises distribution~$\{\rho_*M_{\kappa_*\Omega},\Omega\in\mathbb{S}\}$ is unknown..

At the critical point~$\rho=n$, the family of von Mises distribution~$\{nM_{\kappa_c\Omega},\Omega\in\mathbb{S}\}$ is stable, where~$\kappa_c$ is the unique positive solution of~\eqref{eq-compatibility}. For any initial condition sufficiently close to~$nM_{\kappa_c\Omega}$ for some~$\Omega\in\mathbb{S}$, there exists~$\Omega_\infty$ such that the solution converges exponentially fast to~$nM_{\kappa_c\Omega_\infty}$. However, in this case, the stability of the uniform distribution~$f=n$ is unknown.

As previously, in the special case~$J_{f_0}=0$, the equation reduces to the heat equation and the solution converges to the uniform equilibrium.

\begin{figure}[!h]
\begin{center}
\scalebox{.9}{\input{c_versus_rho_2d_3d.pspdftex}}
\caption{Phase diagram of the model with hysteresis, in dimensions~$2$ and~$3$.}
\label{fig-phase-diagram}
\end{center}
\end{figure}
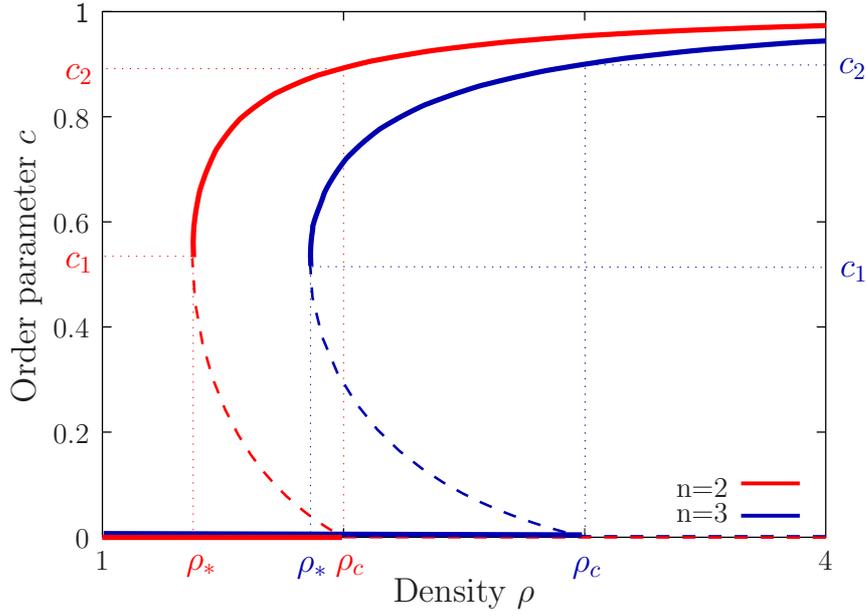

Since~$c(\kappa)$ is an increasing function of~$\kappa$, we can invert this relation~$\kappa\mapsto c(\kappa)$ into~$c\mapsto\kappa(c)$ and express the density~$\rho=\frac{\sigma(\kappa(c))}c$ as a function of~$c$. The result is depicted in Figure~\ref{fig-phase-diagram} for dimension~$2$ or~$3$. With this picture, we recover the phase diagram in a conventional way: the possible order parameters~$c$ for the different equilibria are given as functions of~$\rho$. The dashed lines corresponds to branches of equilibria which are unstable.

\begin{figure}[!h]
\begin{center}
\scalebox{.6}{\input{f_versus_rho_2d_3d.pspdftex}}\hfill\scalebox{.6}{\input{diff_f_versus_rho_2d_3d.pspdftex}}
\caption{Free energy levels of the different equilibria (left), and difference of free energies~$\mathcal F_\kappa-\mathcal F(\rho)$ (right), as functions of the density, in dimensions~$2$ and~$3$.}
\label{fig-free-energy}
\end{center}
\end{figure}
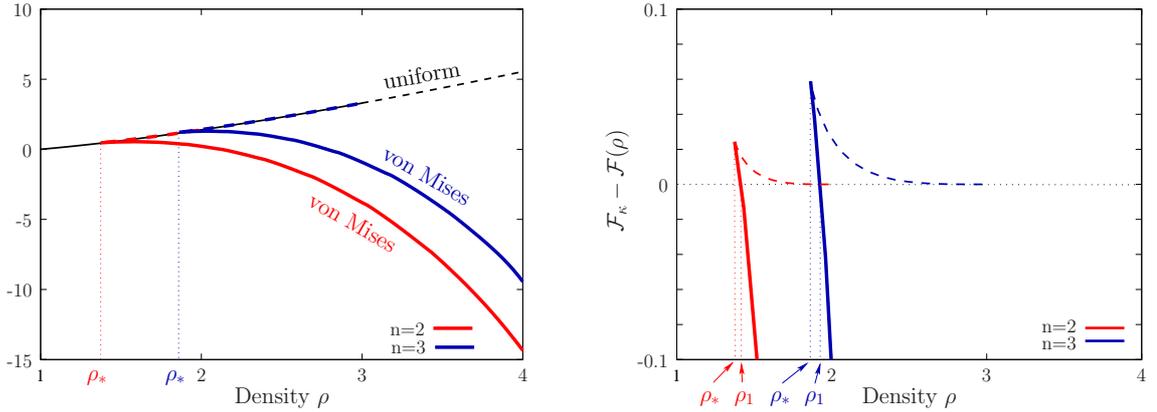

We can also obtain the corresponding diagrams for the free energy and the rates of convergences.
For this particular example, the free energies~$\mathcal F(\rho)$ and~$\mathcal F_\kappa$ (we recall that they correspond respectively to the free energy of the uniform distribution and of a von Mises distribution~$\rho M_{\kappa\Omega}$ for a positive solution~$\kappa$ of the compatibility equation~\eqref{eq-compatibility}, including both stable and unstable branches) are given by

\begin{gather*}
\mathcal F(\rho)=\rho\ln \rho,\\
\begin{split}
\mathcal F_\kappa&=\rho\ln \rho + \langle\rho\ln M_{\kappa\Omega}\rangle_M - \frac12\sigma^2-\frac13\sigma^3\\
&=\rho\ln \rho  - \rho \ln \int e^{\kappa\cos\theta}\d \omega -\frac16(\kappa-\sigma)+\frac23{\sigma\kappa}.
\end{split}
\end{gather*}
The plots of these functions are depicted in dimensions~$2$ and~$3$ are depicted on the left part of Figure~\ref{fig-free-energy}. Since the functions are very close in the figure for some range of interest, we depict the difference~$\mathcal F_\kappa-\mathcal F(\rho)$ in a more appropriate scale, in the right part of Figure~\ref{fig-free-energy}. The dashed lines correspond to unstable branches of equilibria.

We observe that the free energy of the unstable non-isotropic equilibria (in dashed line) is always above that of the uniform distribution. There exist~$\rho_1\in(\rho_*,\rho_c)$ and a corresponding solution~$\kappa_1$ of the compatibility solution~\eqref{eq-compatibility} (with~$\kappa_1>\kappa_*$, corresponding to a stable family of non-isotropic equilibria) such that~$\mathcal F_{\kappa_1}=\mathcal F(\rho_{\mathcal F})$. If~$\rho<\rho_1$, the global minimizer of the free energy is the uniform distribution, while if~$\rho>\rho_1$, then the global minimum is reached for the family of stable von Mises equilibria. The physical relevance of this value is not clear though, as we will see in the numerical illustration of the next subsection.

\begin{figure}[!h]
\begin{center}
\scalebox{.8}{\input{lambda_versus_rho_2d_3d.pspdftex}}
\caption{Rates of convergence to both types of stable equilibria, as functions of the density~$\rho$, in dimensions~$2$ and~$3$.}
\label{fig-rates}
\end{center}
\end{figure}
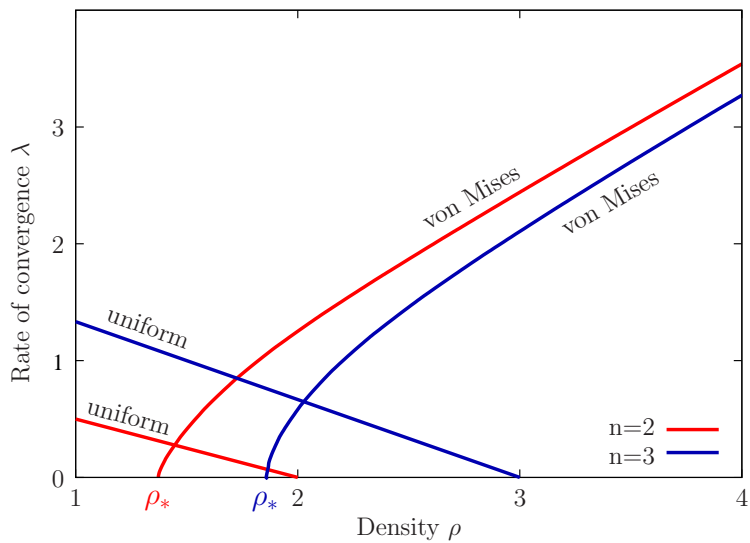

The rates of convergence to the stable equilibria, following Theorems~\ref{thm-uniform} and~\ref{thm-nonisotropic}, are given by
\begin{gather*}
\lambda_0=(n-1)(1-\frac{\rho}n),\text{ for } \rho<\rho_c=n,\\
\lambda_\kappa=\frac{1}{1+\sigma}\Lambda_\kappa(1-(\frac1c-c-\frac{n-1}{\kappa})\sigma(1+2\sigma)),\text{ for } \rho>\rho_*,
\end{gather*}
where~$\lambda_0$ is the rate of convergence to the uniform distribution~$\rho$, and~$\lambda_\kappa$ is the rate of convergence to the stable family of von Mises distributions~$\rho M_{\kappa\Omega}$, where~$\kappa$ is the unique solution of the compatibility condition~\eqref{eq-compatibility} such that~$\kappa>\kappa_*$. Details for the numerical computation of the Poincaré constant~$\Lambda_\kappa$ are given in the appendix of~\cite{degond2012macroscopic}. The computations in dimensions~$2$ and~$3$ are depicted in Figure~\ref{fig-rates}.

\subsection{Numerical illustrations of the hysteresis phenomenon}

In order to highlight the role of the density~$\rho$ as the key parameter for this phase transition, we introduce the probability density function~$\widetilde f=\frac{f}{\rho}$ and we get
\begin{equation}
\label{eq-KFP-rho}
\partial_t\widetilde f = \tau(\rho|J_{\widetilde f}|)\Delta_\omega\widetilde f - \nu(\rho|J_{\widetilde f}|)\nabla_\omega\cdot(\widetilde f\nabla_\omega(\Omega_{\widetilde f}\cdot\omega)).
\end{equation}

When~$\rho$ is constant, this equation is equivalent to~\eqref{FP-eq}. We now consider~$\rho$ as a parameter varying slowly with time (compared to the time scale of the convergence to equilibrium, see Figure~\ref{fig-rates}). If this parameter starts from a value~$\rho<\rho_*$, and increases slowly, the only stable distribution is initially the uniform distribution~$\widetilde f=1$, and it remains stable. So we expect that the solution stays close to it, until~$\rho$ reaches the critical value~$\rho_c$. For~$\rho>\rho_c$, the only stable equilibria are the von Mises distributions, and the solution converges to one of these equilibria. The order parameter defined as~$c(\widetilde f)=|J_{\widetilde f}|$, then jumps from~$0$ to~$c_c =c(\kappa_c)$. If then the density~$\rho$ is further decreased slowly, the solution stays close to a von Mises distribution, and the order parameter slowly decreases, until~$\rho$ reaches~$\rho_*$ back. For~$\rho<\rho_*$, the only stable equilibrium is the uniform distribution, and the concentration parameter jumps from~$c_*=c(\kappa_*)$ to~$0$. This is a hysteresis phenomenon: the concentration parameter describes an oriented loop called hysteresis loop.

Let us now present some numerical simulations of the system~\eqref{eq-KFP-rho} in dimension~$n=2$. We start with a initial condition which is a small perturbation of the uniform distribution, and we take~$\rho=1.75-0.75\cos(\frac{\pi}Tt)$, with~$T=500$. We use a standard central finite different scheme (with~$100$ discretization points), implicit in time (with a time step of~$0.01$).
The only problem with this approach is that the solution converges so strongly to the uniform distribution for~$\rho<\rho_c$, so after passing~$\rho_c$, the linear rate of explosion for~$J_{\widetilde f}$ is given by~$e^{(\frac{\rho}{\rho_c}-1)t}$, and is very slow when~$\rho$ is close to~$\rho_c$. So since $J_{\widetilde f}$ is initially very small when passing the threshold~$\rho=\rho_c$ we would have to wait extremely long in order to see the convergence to the stable von Mises distribution. To overcome this problem, we adding a threshold~$\varepsilon$ and strengthen~$|J_{\widetilde f}|$ when~$\|\widetilde f-1\|_\infty\leqslant\varepsilon$, by
\[
\widetilde f\rightsquigarrow \widetilde f + \max(0,\varepsilon-\|\widetilde f-1\|_\infty)\,\Omega_{\widetilde f}\cdot\omega.
\]
We note that this transformation that we still have~$\|\widetilde f-1\|_\infty\leqslant\varepsilon$ if it was the case before applying the transformation.

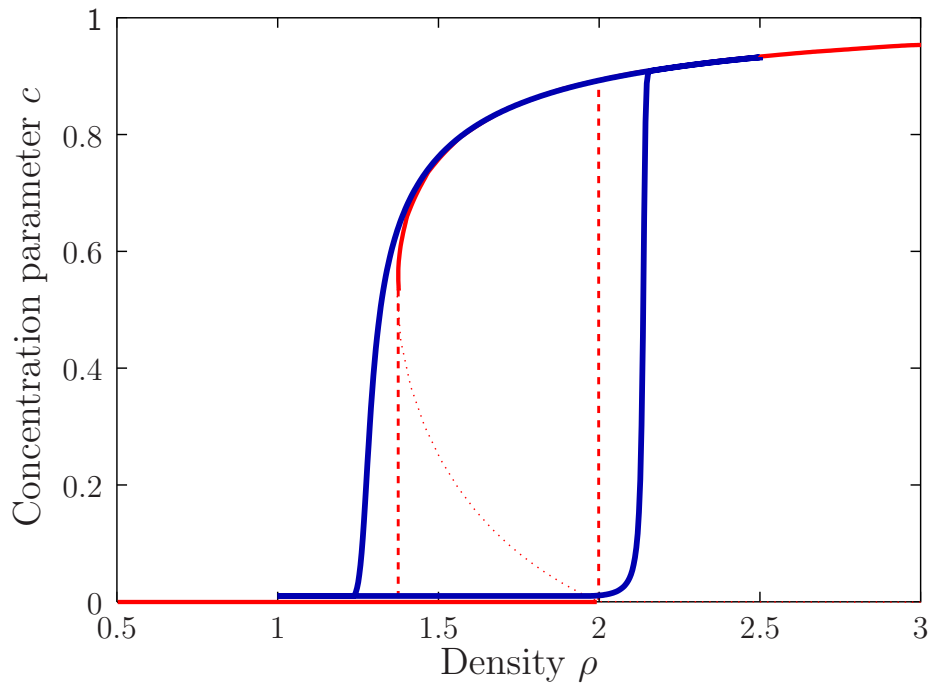
\begin{figure}[!h]
\begin{center}
\input{c_kinetic_500.pspdftex}
\caption{Hysteresis loop for the concentration parameter~$c$ in a numerical simulation of the kinetic equation~\eqref{eq-KFP-rho}, with time varying~$\rho$, in dimension~$2$. The red curve is the theoretical curve, the blue one corresponds to the simulation.}
\label{fig-numerics-kinetic-2d}
\end{center}
\end{figure}

Figure~\ref{fig-numerics-kinetic-2d} depicts the result of a numerical simulation with a threshold~$\varepsilon=0.02$. We clearly see this hysteresis cycle, which agrees very well with the theoretical diagram. The jumps at~$\rho=\rho_*$ and~$\rho=\rho_c$ are closer to the theoretical jumps when~$T$ is very large.
We were not able to see any numerical significance of the value~$\rho_1$ (for which uniform and non-isotropic distributions have the same free energy) in all these numerical simulations. In particular,~$\rho_1$ is close to~$\rho_*$ (see Figure~\ref{fig-free-energy}), so in most of the cases where both uniform and non-isotropic distribution are stable, the uniform distribution is not the global minimizer of the free energy, but in practical, meta-stability is very strong, and the solution still converges to the uniform distribution.

\section{Conclusion}
In this note, we have given the summary of strong results on the stability and instability of the equilibrium states of the modified Smoluchowski equation~\eqref{FP-eq}. This allows to have a precise description of the dynamics of the solution when time goes to infinity: it converges exponentially fast to a fixed equilibrium, with explicit formulas for the rates of convergence. We have also exhibited a specific example in which we observe a first order phase transition with a hysteresis loop (in contrast with the second order phase transition of the original Smoluchowski equation with dipolar potential~\cite{frouvelle2012dynamics}). The details of the proofs will be found in a longer paper~\cite{degond2013phase} as well as numerical comparisons between the particle and kinetic models to confirm that the hysteresis is really intrinsic to the system and not simply an artifact of the kinetic modeling.

\end{document}

%% file: shapes_c_sigma_versus_kappa.pspdftex
\begin{picture}(0,0)%
\includegraphics{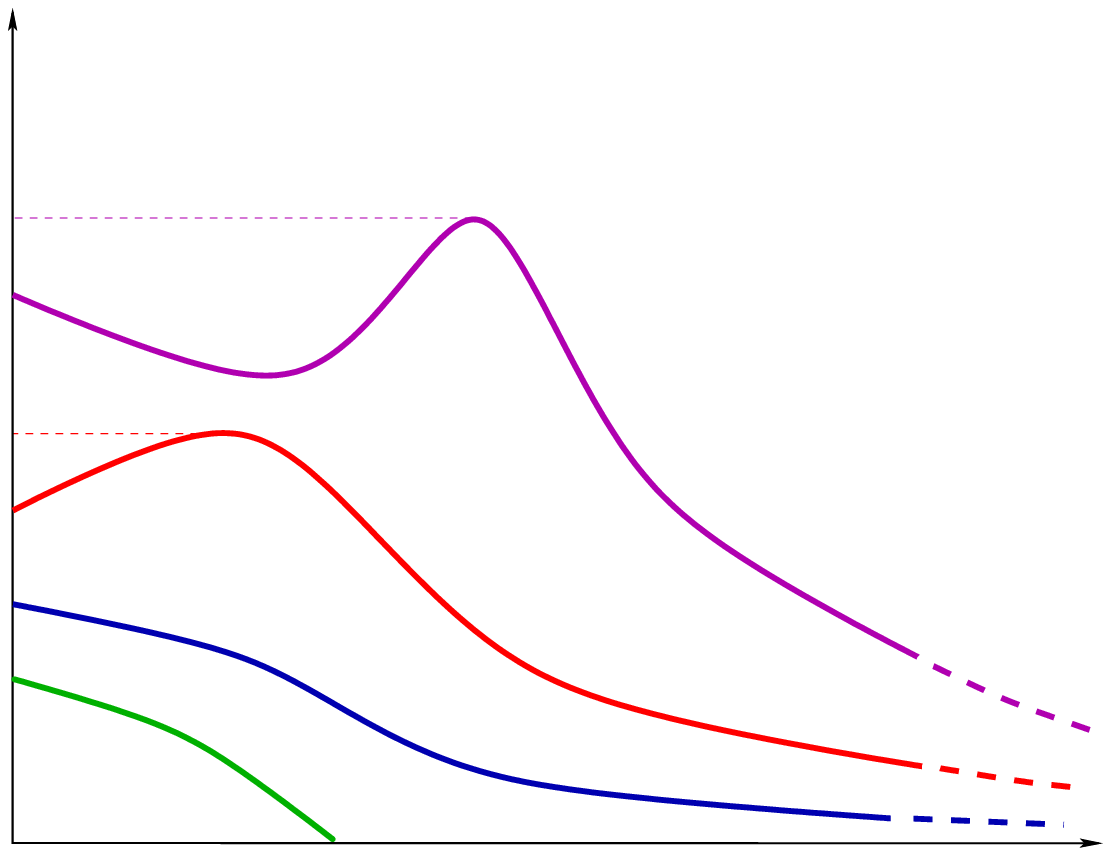}%
\end{picture}%
\setlength{\unitlength}{2368sp}%
\begingroup\makeatletter\ifx\SetFigFont\undefined%
\gdef\SetFigFont#1#2#3#4#5{%
  \reset@font\fontsize{#1}{#2pt}%
  \fontfamily{#3}\fontseries{#4}\fontshape{#5}%
  \selectfont}%
\fi\endgroup%
\begin{picture}(9074,7207)(1691,-7276)
\put(2026,-7136){\makebox(0,0)[b]{\smash{{\SetFigFont{12}{14.4}{\familydefault}{\mddefault}{\updefault}0}}}}
\put(1876,-6886){\makebox(0,0)[rb]{\smash{{\SetFigFont{12}{14.4}{\familydefault}{\mddefault}{\updefault}0}}}}
\put(4651,-7111){\makebox(0,0)[b]{\smash{{\SetFigFont{14}{16.8}{\familydefault}{\mddefault}{\updefault}{\color[rgb]{0,.69,0}$\kappa_{max}$}%
}}}}
\put(1951,-5611){\makebox(0,0)[rb]{\smash{{\SetFigFont{14}{16.8}{\familydefault}{\mddefault}{\updefault}{\color[rgb]{0,.69,0}$\frac1{\rho_c}=\frac1{\rho_*}$}%
}}}}
\put(1951,-5011){\makebox(0,0)[rb]{\smash{{\SetFigFont{14}{16.8}{\familydefault}{\mddefault}{\updefault}{\color[rgb]{0,0,.69}$\frac1{\rho_c}=\frac1{\rho_*}$}%
}}}}
\put(1951,-4276){\makebox(0,0)[rb]{\smash{{\SetFigFont{14}{16.8}{\familydefault}{\mddefault}{\updefault}{\color[rgb]{1,0,0}$\frac1{\rho_c}$}%
}}}}
\put(1951,-2506){\makebox(0,0)[rb]{\smash{{\SetFigFont{14}{16.8}{\familydefault}{\mddefault}{\updefault}{\color[rgb]{.69,0,.69}$\frac1{\rho_c}$}%
}}}}
\put(1966,-1921){\makebox(0,0)[rb]{\smash{{\SetFigFont{14}{16.8}{\familydefault}{\mddefault}{\updefault}{\color[rgb]{.69,0,.69}$\frac1{\rho_*}$}%
}}}}
\put(1954,-3643){\makebox(0,0)[rb]{\smash{{\SetFigFont{14}{16.8}{\familydefault}{\mddefault}{\updefault}{\color[rgb]{1,0,0}$\frac1{\rho_*}$}%
}}}}
\put(10271,-7144){\makebox(0,0)[b]{\smash{{\SetFigFont{14}{16.8}{\familydefault}{\mddefault}{\updefault}$\kappa$}}}}
\put(1918,-694){\makebox(0,0)[rb]{\smash{{\SetFigFont{14}{16.8}{\familydefault}{\mddefault}{\updefault}$\frac{c(\kappa)}{\sigma(\kappa)}$}}}}
\end{picture}%

%% file: c_sigma_versus_kappa.pspdftex
\begin{picture}(0,0)%
\includegraphics{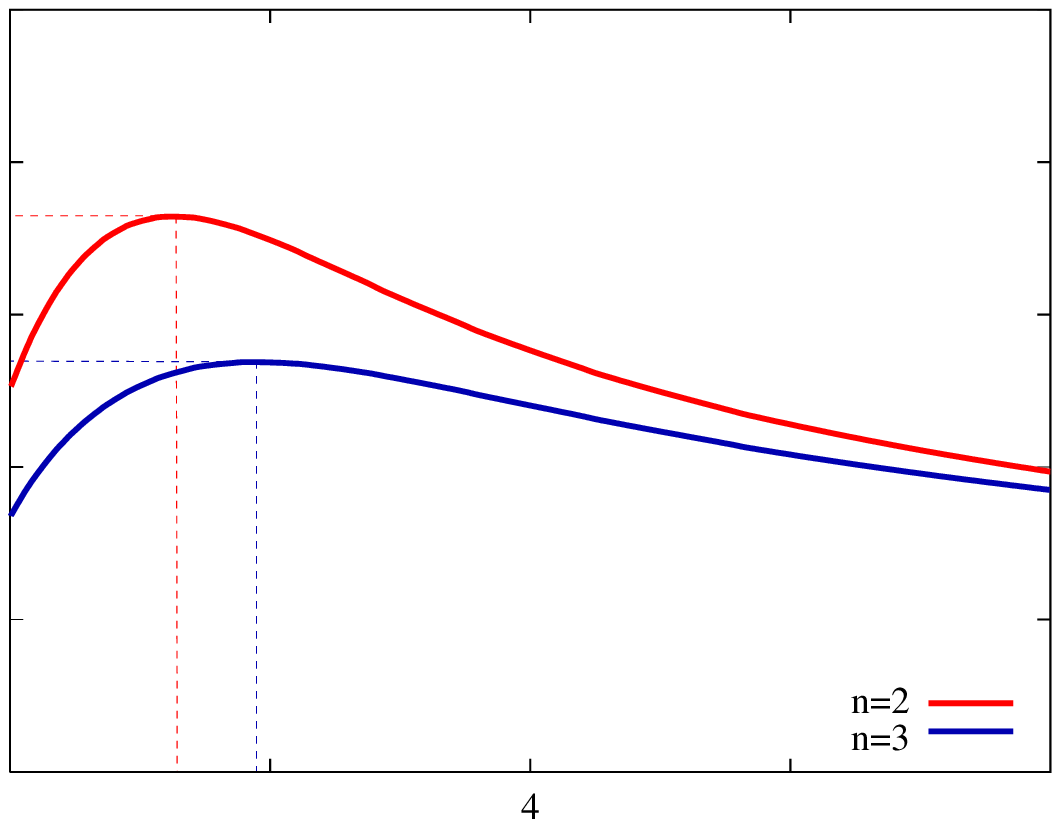}%
\end{picture}%
\setlength{\unitlength}{2368sp}%
\begingroup\makeatletter\ifx\SetFigFont\undefined%
\gdef\SetFigFont#1#2#3#4#5{%
  \reset@font\fontsize{#1}{#2pt}%
  \fontfamily{#3}\fontseries{#4}\fontshape{#5}%
  \selectfont}%
\fi\endgroup%
\begin{picture}(9158,7141)(1291,-7688)
\put(1876,-3227){\makebox(0,0)[rb]{\smash{{\SetFigFont{12}{14.4}{\familydefault}{\mddefault}{\updefault}0.6}}}}
\put(1876,-2007){\makebox(0,0)[rb]{\smash{{\SetFigFont{12}{14.4}{\familydefault}{\mddefault}{\updefault}0.8}}}}
\put(1876,-787){\makebox(0,0)[rb]{\smash{{\SetFigFont{12}{14.4}{\familydefault}{\mddefault}{\updefault}1}}}}
\put(1876,-6886){\makebox(0,0)[rb]{\smash{{\SetFigFont{12}{14.4}{\familydefault}{\mddefault}{\updefault}0}}}}
\put(1876,-5666){\makebox(0,0)[rb]{\smash{{\SetFigFont{12}{14.4}{\familydefault}{\mddefault}{\updefault}0.2}}}}
\put(1876,-4446){\makebox(0,0)[rb]{\smash{{\SetFigFont{12}{14.4}{\familydefault}{\mddefault}{\updefault}0.4}}}}
\put(2026,-7136){\makebox(0,0)[b]{\smash{{\SetFigFont{12}{14.4}{\familydefault}{\mddefault}{\updefault}0}}}}
\put(10350,-7136){\makebox(0,0)[b]{\smash{{\SetFigFont{12}{14.4}{\familydefault}{\mddefault}{\updefault}8}}}}
\put(1895,-2425){\makebox(0,0)[rb]{\smash{{\SetFigFont{14}{16.8}{\familydefault}{\mddefault}{\updefault}{\color[rgb]{1,0,0}$\frac1{\rho_*}$}%
}}}}
\put(4010,-7179){\makebox(0,0)[b]{\smash{{\SetFigFont{14}{16.8}{\familydefault}{\mddefault}{\updefault}{\color[rgb]{0,0,.69}$\kappa_*$}%
}}}}
\put(3410,-7179){\makebox(0,0)[b]{\smash{{\SetFigFont{14}{16.8}{\familydefault}{\mddefault}{\updefault}{\color[rgb]{1,0,0}$\kappa_*$}%
}}}}
\put(6158,-7556){\makebox(0,0)[b]{\smash{{\SetFigFont{14}{16.8}{\familydefault}{\mddefault}{\updefault}$\kappa$}}}}
\put(1899,-3653){\makebox(0,0)[rb]{\smash{{\SetFigFont{14}{16.8}{\familydefault}{\mddefault}{\updefault}{\color[rgb]{0,0,.69}$\frac1{\rho_*}$}%
}}}}
\put(1306,-3886){\makebox(0,0)[rb]{\smash{{\SetFigFont{14}{16.8}{\familydefault}{\mddefault}{\updefault}$\frac{c(\kappa)}{\sigma(\kappa)}$}}}}
\end{picture}%

%% file: c_versus_rho_2d_3d.pspdftex
\begin{picture}(0,0)%
\includegraphics{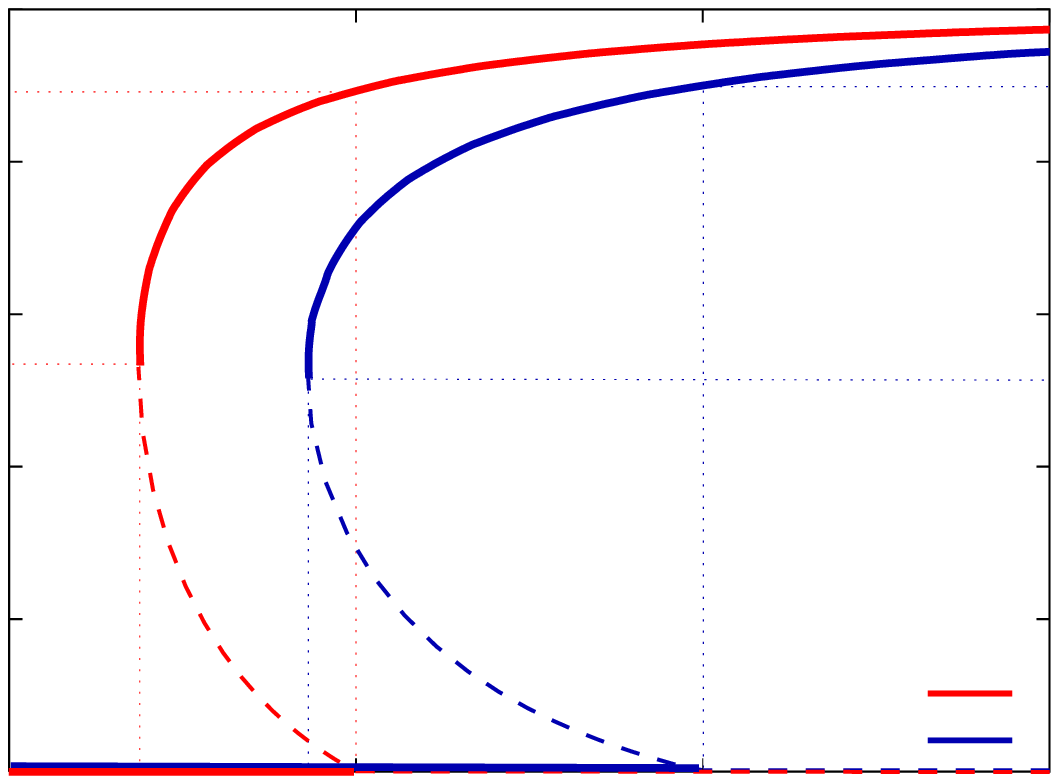}%
\end{picture}%
\setlength{\unitlength}{2368sp}%
\begingroup\makeatletter\ifx\SetFigFont\undefined%
\gdef\SetFigFont#1#2#3#4#5{%
  \reset@font\fontsize{#1}{#2pt}%
  \fontfamily{#3}\fontseries{#4}\fontshape{#5}%
  \selectfont}%
\fi\endgroup%
\begin{picture}(9601,7096)(930,-7643)
\put(1876,-787){\makebox(0,0)[rb]{\smash{{\SetFigFont{12}{14.4}{\sfdefault}{\mddefault}{\updefault}1}}}}
\put(2026,-7136){\makebox(0,0)[b]{\smash{{\SetFigFont{12}{14.4}{\familydefault}{\mddefault}{\updefault}1}}}}
\put(10350,-7136){\makebox(0,0)[b]{\smash{{\SetFigFont{12}{14.4}{\familydefault}{\mddefault}{\updefault}4}}}}
\put(9225,-6586){\makebox(0,0)[rb]{\smash{{\SetFigFont{12}{14.4}{\familydefault}{\mddefault}{\updefault}n=3}}}}
\put(9225,-6286){\makebox(0,0)[rb]{\smash{{\SetFigFont{12}{14.4}{\familydefault}{\mddefault}{\updefault}n=2}}}}
\put(1876,-5666){\makebox(0,0)[rb]{\smash{{\SetFigFont{12}{14.4}{\familydefault}{\mddefault}{\updefault}0.2}}}}
\put(1876,-6886){\makebox(0,0)[rb]{\smash{{\SetFigFont{12}{14.4}{\familydefault}{\mddefault}{\updefault}0}}}}
\put(1876,-4446){\makebox(0,0)[rb]{\smash{{\SetFigFont{12}{14.4}{\familydefault}{\mddefault}{\updefault}0.4}}}}
\put(1876,-3227){\makebox(0,0)[rb]{\smash{{\SetFigFont{12}{14.4}{\familydefault}{\mddefault}{\updefault}0.6}}}}
\put(1876,-2007){\makebox(0,0)[rb]{\smash{{\SetFigFont{12}{14.4}{\familydefault}{\mddefault}{\updefault}0.8}}}}
\put(4441,-7156){\makebox(0,0)[b]{\smash{{\SetFigFont{14}{16.8}{\familydefault}{\mddefault}{\updefault}{\color[rgb]{0,0,.69}$\rho_*$}%
}}}}
\put(3166,-7141){\makebox(0,0)[b]{\smash{{\SetFigFont{14}{16.8}{\familydefault}{\mddefault}{\updefault}{\color[rgb]{1,0,0}$\rho_*$}%
}}}}
\put(7606,-7156){\makebox(0,0)[b]{\smash{{\SetFigFont{14}{16.8}{\familydefault}{\mddefault}{\updefault}{\color[rgb]{0,0,.69}$\rho_c$}%
}}}}
\put(4876,-7141){\makebox(0,0)[b]{\smash{{\SetFigFont{14}{16.8}{\familydefault}{\mddefault}{\updefault}{\color[rgb]{1,0,0}$\rho_c$}%
}}}}
\put(1921,-3631){\makebox(0,0)[rb]{\smash{{\SetFigFont{14}{16.8}{\familydefault}{\mddefault}{\updefault}{\color[rgb]{1,0,0}$c_1$}%
}}}}
\put(1921,-1441){\makebox(0,0)[rb]{\smash{{\SetFigFont{14}{16.8}{\familydefault}{\mddefault}{\updefault}{\color[rgb]{1,0,0}$c_2$}%
}}}}
\put(10501,-1381){\makebox(0,0)[lb]{\smash{{\SetFigFont{14}{16.8}{\familydefault}{\mddefault}{\updefault}{\color[rgb]{0,0,.69}$c_2$}%
}}}}
\put(10516,-3736){\makebox(0,0)[lb]{\smash{{\SetFigFont{14}{16.8}{\familydefault}{\mddefault}{\updefault}{\color[rgb]{0,0,.69}$c_1$}%
}}}}
\put(1257,-3619){\rotatebox{90.0}{\makebox(0,0)[b]{\smash{{\SetFigFont{14}{16.8}{\familydefault}{\mddefault}{\updefault}Order parameter $c$}}}}}
\put(6188,-7511){\makebox(0,0)[b]{\smash{{\SetFigFont{14}{16.8}{\familydefault}{\mddefault}{\updefault}Density $\rho$}}}}
\end{picture}%

%% file: f_versus_rho_2d_3d.pspdftex
\begin{picture}(0,0)%
\includegraphics{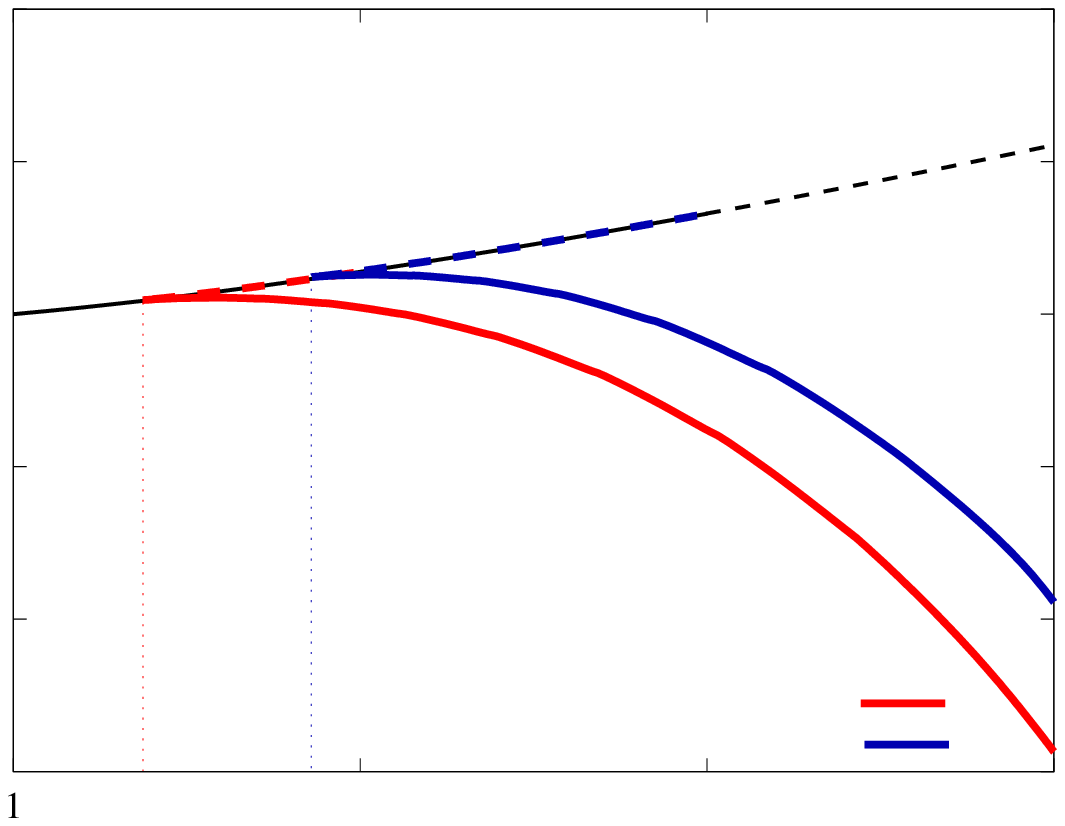}%
\end{picture}%
\setlength{\unitlength}{2368sp}%
\begingroup\makeatletter\ifx\SetFigFont\undefined%
\gdef\SetFigFont#1#2#3#4#5{%
  \reset@font\fontsize{#1}{#2pt}%
  \fontfamily{#3}\fontseries{#4}\fontshape{#5}%
  \selectfont}%
\fi\endgroup%
\begin{picture}(9035,7096)(1414,-7643)
\put(4801,-7136){\makebox(0,0)[b]{\smash{{\SetFigFont{12}{14.4}{\familydefault}{\mddefault}{\updefault}2}}}}
\put(7575,-7136){\makebox(0,0)[b]{\smash{{\SetFigFont{12}{14.4}{\familydefault}{\mddefault}{\updefault}3}}}}
\put(10350,-7136){\makebox(0,0)[b]{\smash{{\SetFigFont{12}{14.4}{\familydefault}{\mddefault}{\updefault}4}}}}
\put(1876,-6886){\makebox(0,0)[rb]{\smash{{\SetFigFont{12}{14.4}{\familydefault}{\mddefault}{\updefault}-15}}}}
\put(1876,-2007){\makebox(0,0)[rb]{\smash{{\SetFigFont{12}{14.4}{\familydefault}{\mddefault}{\updefault}5}}}}
\put(1876,-5666){\makebox(0,0)[rb]{\smash{{\SetFigFont{12}{14.4}{\familydefault}{\mddefault}{\updefault}-10}}}}
\put(1876,-4446){\makebox(0,0)[rb]{\smash{{\SetFigFont{12}{14.4}{\familydefault}{\mddefault}{\updefault}-5}}}}
\put(1876,-3227){\makebox(0,0)[rb]{\smash{{\SetFigFont{12}{14.4}{\familydefault}{\mddefault}{\updefault}0}}}}
\put(1876,-787){\makebox(0,0)[rb]{\smash{{\SetFigFont{12}{14.4}{\familydefault}{\mddefault}{\updefault}10}}}}
\put(6188,-7511){\makebox(0,0)[b]{\smash{{\SetFigFont{14}{16.8}{\familydefault}{\mddefault}{\updefault}Density $\rho$}}}}
\put(8671,-6646){\makebox(0,0)[rb]{\smash{{\SetFigFont{12}{14.4}{\familydefault}{\mddefault}{\updefault}n=3}}}}
\put(8656,-6331){\makebox(0,0)[rb]{\smash{{\SetFigFont{12}{14.4}{\familydefault}{\mddefault}{\updefault}n=2}}}}
\put(3001,-7111){\makebox(0,0)[b]{\smash{{\SetFigFont{14}{16.8}{\familydefault}{\mddefault}{\updefault}{\color[rgb]{1,0,0}$\rho_*$}%
}}}}
\put(4351,-7111){\makebox(0,0)[b]{\smash{{\SetFigFont{14}{16.8}{\familydefault}{\mddefault}{\updefault}{\color[rgb]{0,0,.69}$\rho_*$}%
}}}}
\put(7351,-4486){\rotatebox{330.0}{\makebox(0,0)[b]{\smash{{\SetFigFont{14}{16.8}{\familydefault}{\mddefault}{\updefault}{\color[rgb]{1,0,0}von Mises}%
}}}}}
\put(8626,-3661){\rotatebox{330.0}{\makebox(0,0)[b]{\smash{{\SetFigFont{14}{16.8}{\familydefault}{\mddefault}{\updefault}{\color[rgb]{0,0,.69}von Mises}%
}}}}}
\put(8626,-1936){\rotatebox{10.0}{\makebox(0,0)[b]{\smash{{\SetFigFont{14}{16.8}{\familydefault}{\mddefault}{\updefault}{\color[rgb]{0,0,0}uniform}%
}}}}}
\end{picture}%

%% file: diff_f_versus_rho_2d_3d.pspdftex
\begin{picture}(0,0)%
\includegraphics{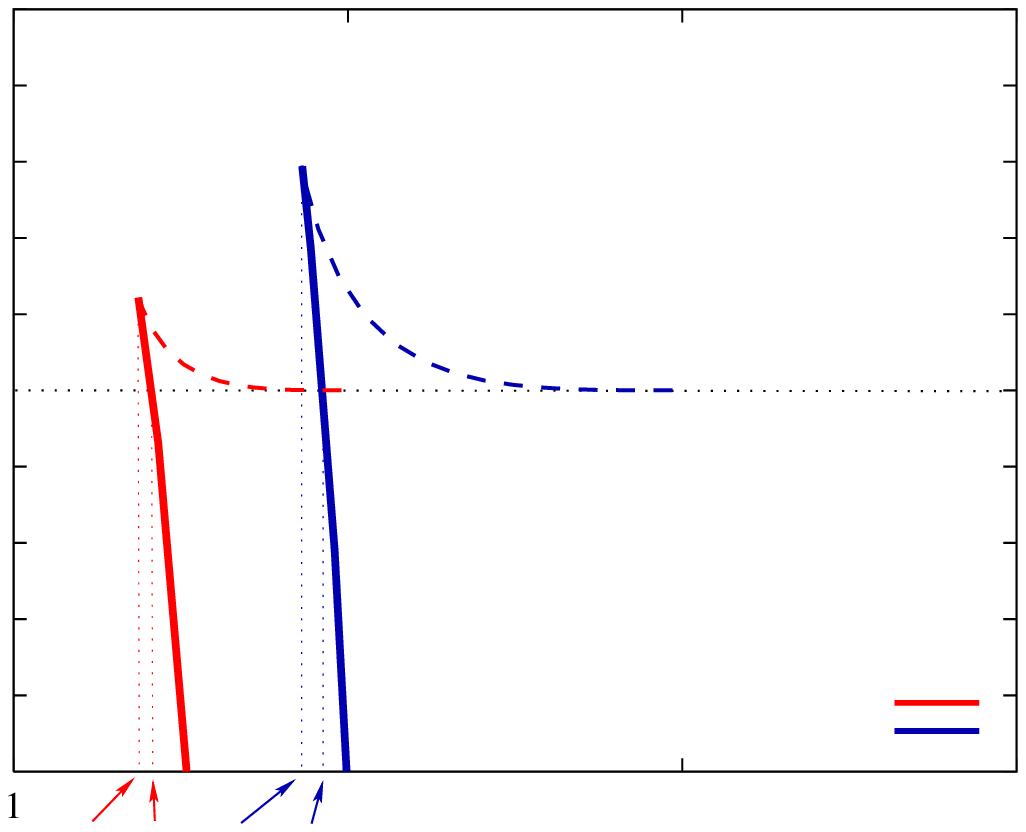}%
\end{picture}%
\setlength{\unitlength}{2368sp}%
\begingroup\makeatletter\ifx\SetFigFont\undefined%
\gdef\SetFigFont#1#2#3#4#5{%
  \reset@font\fontsize{#1}{#2pt}%
  \fontfamily{#3}\fontseries{#4}\fontshape{#5}%
  \selectfont}%
\fi\endgroup%
\begin{picture}(9375,7096)(1074,-7643)
\put(10350,-7136){\makebox(0,0)[b]{\smash{{\SetFigFont{12}{14.4}{\familydefault}{\mddefault}{\updefault}4}}}}
\put(7675,-7136){\makebox(0,0)[b]{\smash{{\SetFigFont{12}{14.4}{\familydefault}{\mddefault}{\updefault}3}}}}
\put(5001,-7136){\makebox(0,0)[b]{\smash{{\SetFigFont{12}{14.4}{\familydefault}{\mddefault}{\updefault}2}}}}
\put(6338,-7511){\makebox(0,0)[b]{\smash{{\SetFigFont{14}{16.8}{\familydefault}{\mddefault}{\updefault}Density $\rho$}}}}
\put(2176,-6886){\makebox(0,0)[rb]{\smash{{\SetFigFont{12}{14.4}{\familydefault}{\mddefault}{\updefault}-0.1}}}}
\put(2176,-787){\makebox(0,0)[rb]{\smash{{\SetFigFont{12}{14.4}{\familydefault}{\mddefault}{\updefault}0.1}}}}
\put(9225,-6586){\makebox(0,0)[rb]{\smash{{\SetFigFont{12}{14.4}{\familydefault}{\mddefault}{\updefault}n=3}}}}
\put(2176,-3836){\makebox(0,0)[rb]{\smash{{\SetFigFont{12}{14.4}{\familydefault}{\mddefault}{\updefault}0}}}}
\put(1401,-3639){\rotatebox{90.0}{\makebox(0,0)[b]{\smash{{\SetFigFont{14}{16.8}{\familydefault}{\mddefault}{\updefault}$\mathcal{F}_\kappa-\mathcal{F}(\rho)$}}}}}
\put(4113,-7479){\makebox(0,0)[b]{\smash{{\SetFigFont{14}{16.8}{\familydefault}{\mddefault}{\updefault}{\color[rgb]{0,0,.69}$\rho_*$}%
}}}}
\put(4707,-7479){\makebox(0,0)[b]{\smash{{\SetFigFont{14}{16.8}{\familydefault}{\mddefault}{\updefault}{\color[rgb]{0,0,.69}$\rho_1$}%
}}}}
\put(2902,-7483){\makebox(0,0)[b]{\smash{{\SetFigFont{14}{16.8}{\familydefault}{\mddefault}{\updefault}{\color[rgb]{1,0,0}$\rho_*$}%
}}}}
\put(3497,-7483){\makebox(0,0)[b]{\smash{{\SetFigFont{14}{16.8}{\familydefault}{\mddefault}{\updefault}{\color[rgb]{1,0,0}$\rho_1$}%
}}}}
\put(9225,-6286){\makebox(0,0)[rb]{\smash{{\SetFigFont{12}{14.4}{\familydefault}{\mddefault}{\updefault}n=2}}}}
\end{picture}%

%% file: lambda_versus_rho_2d_3d.pspdftex
\begin{picture}(0,0)%
\includegraphics{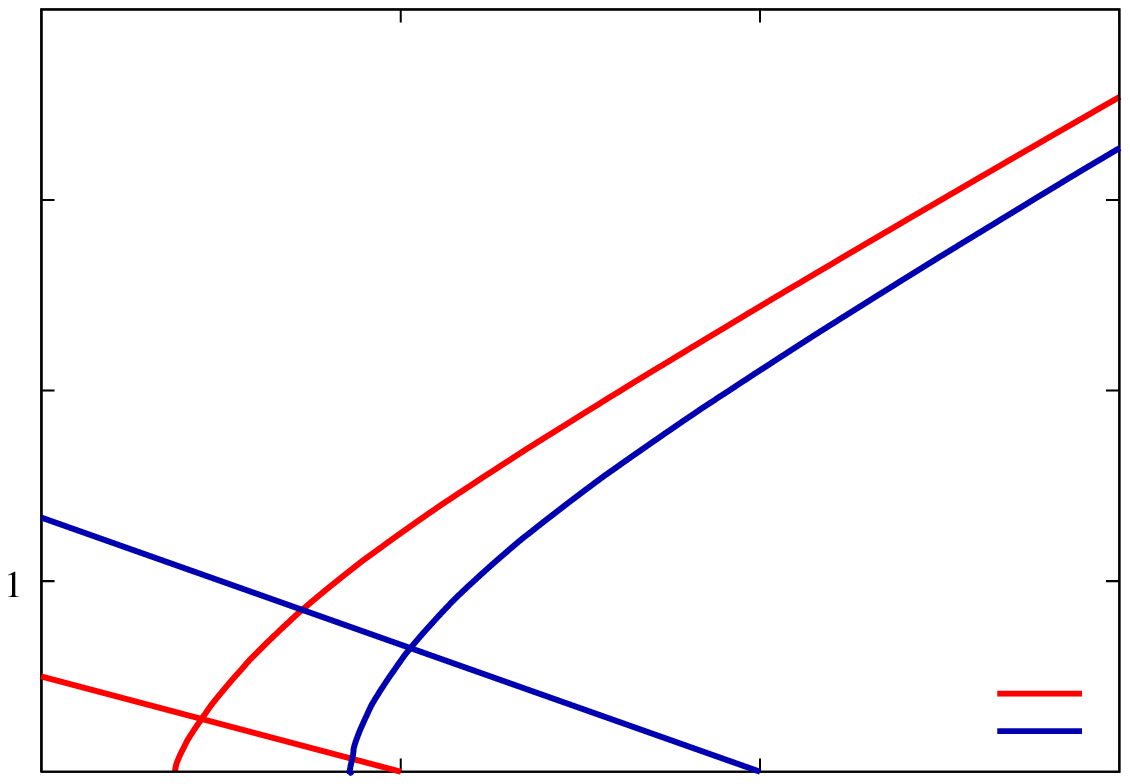}%
\end{picture}%
\setlength{\unitlength}{2368sp}%
\begingroup\makeatletter\ifx\SetFigFont\undefined%
\gdef\SetFigFont#1#2#3#4#5{%
  \reset@font\fontsize{#1}{#2pt}%
  \fontfamily{#3}\fontseries{#4}\fontshape{#5}%
  \selectfont}%
\fi\endgroup%
\begin{picture}(9590,6980)(859,-7620)
\put(6038,-7511){\makebox(0,0)[b]{\smash{{\SetFigFont{12}{14.4}{\familydefault}{\mddefault}{\updefault}Density $\rho$}}}}
\put(7475,-7136){\makebox(0,0)[b]{\smash{{\SetFigFont{12}{14.4}{\familydefault}{\mddefault}{\updefault}3}}}}
\put(10350,-7136){\makebox(0,0)[b]{\smash{{\SetFigFont{12}{14.4}{\familydefault}{\mddefault}{\updefault}4}}}}
\put(4601,-7136){\makebox(0,0)[b]{\smash{{\SetFigFont{12}{14.4}{\familydefault}{\mddefault}{\updefault}2}}}}
\put(1726,-7136){\makebox(0,0)[b]{\smash{{\SetFigFont{12}{14.4}{\familydefault}{\mddefault}{\updefault}1}}}}
\put(1576,-6886){\makebox(0,0)[rb]{\smash{{\SetFigFont{12}{14.4}{\familydefault}{\mddefault}{\updefault}0}}}}
\put(1576,-3836){\makebox(0,0)[rb]{\smash{{\SetFigFont{12}{14.4}{\familydefault}{\mddefault}{\updefault}2}}}}
\put(1576,-2312){\makebox(0,0)[rb]{\smash{{\SetFigFont{12}{14.4}{\familydefault}{\mddefault}{\updefault}3}}}}
\put(1126,-3961){\rotatebox{90.0}{\makebox(0,0)[b]{\smash{{\SetFigFont{12}{14.4}{\familydefault}{\mddefault}{\updefault}Rate of convergence $\lambda$}}}}}
\put(3151,-5086){\rotatebox{340.0}{\makebox(0,0)[rb]{\smash{{\SetFigFont{12}{14.4}{\familydefault}{\mddefault}{\updefault}uniform}}}}}
\put(2926,-6211){\rotatebox{345.0}{\makebox(0,0)[rb]{\smash{{\SetFigFont{12}{14.4}{\familydefault}{\mddefault}{\updefault}uniform}}}}}
\put(7501,-2836){\rotatebox{30.0}{\makebox(0,0)[rb]{\smash{{\SetFigFont{12}{14.4}{\familydefault}{\mddefault}{\updefault}von Mises}}}}}
\put(9301,-2911){\rotatebox{30.0}{\makebox(0,0)[rb]{\smash{{\SetFigFont{12}{14.4}{\familydefault}{\mddefault}{\updefault}von Mises}}}}}
\put(9225,-6211){\makebox(0,0)[rb]{\smash{{\SetFigFont{12}{14.4}{\familydefault}{\mddefault}{\updefault}n=2}}}}
\put(9225,-6586){\makebox(0,0)[rb]{\smash{{\SetFigFont{12}{14.4}{\familydefault}{\mddefault}{\updefault}n=3}}}}
\put(2791,-7112){\makebox(0,0)[b]{\smash{{\SetFigFont{14}{16.8}{\familydefault}{\mddefault}{\updefault}{\color[rgb]{1,0,0}$\rho_*$}%
}}}}
\put(4181,-7124){\makebox(0,0)[b]{\smash{{\SetFigFont{14}{16.8}{\familydefault}{\mddefault}{\updefault}{\color[rgb]{0,0,.69}$\rho_*$}%
}}}}
\end{picture}%

%% file: c_kinetic_500.pspdftex
\begin{picture}(0,0)%
\includegraphics{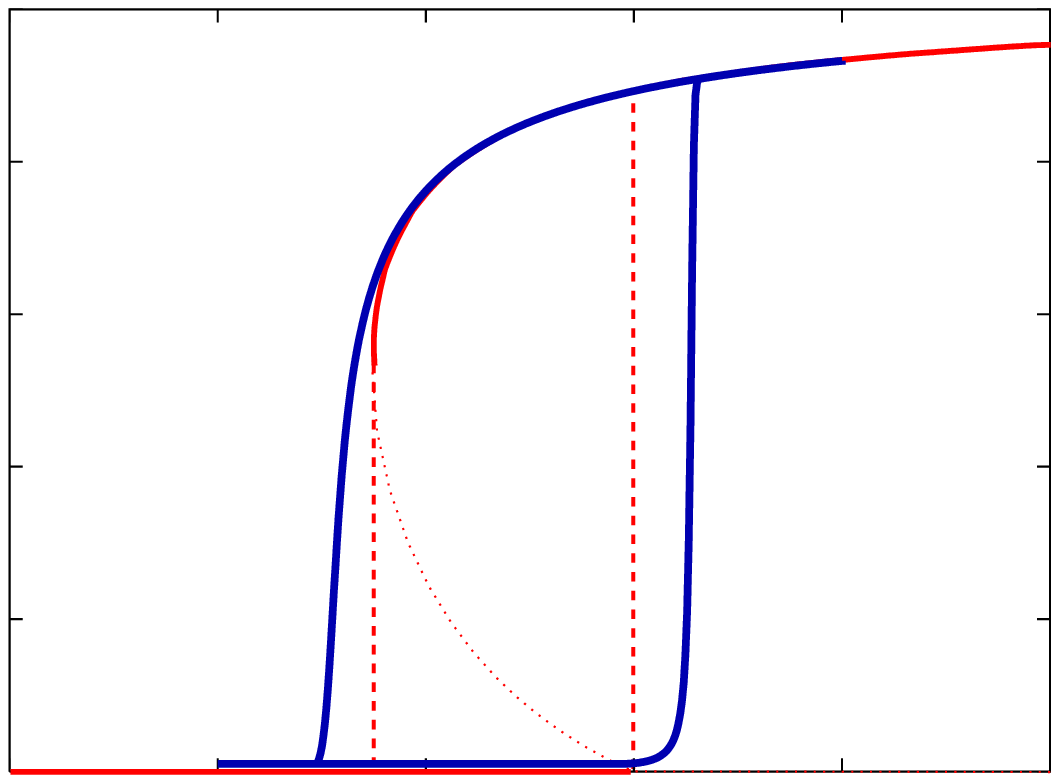}%
\end{picture}%
\setlength{\unitlength}{2368sp}%
\begingroup\makeatletter\ifx\SetFigFont\undefined%
\gdef\SetFigFont#1#2#3#4#5{%
  \reset@font\fontsize{#1}{#2pt}%
  \fontfamily{#3}\fontseries{#4}\fontshape{#5}%
  \selectfont}%
\fi\endgroup%
\begin{picture}(9549,7096)(900,-7643)
\put(1876,-787){\makebox(0,0)[rb]{\smash{{\SetFigFont{12}{14.4}{\sfdefault}{\mddefault}{\updefault}1}}}}
\put(6188,-7511){\makebox(0,0)[b]{\smash{{\SetFigFont{14}{16.8}{\familydefault}{\mddefault}{\updefault}Density $\rho$}}}}
\put(10350,-7136){\makebox(0,0)[b]{\smash{{\SetFigFont{12}{14.4}{\familydefault}{\mddefault}{\updefault}3}}}}
\put(8685,-7136){\makebox(0,0)[b]{\smash{{\SetFigFont{12}{14.4}{\familydefault}{\mddefault}{\updefault}2.5}}}}
\put(7020,-7136){\makebox(0,0)[b]{\smash{{\SetFigFont{12}{14.4}{\familydefault}{\mddefault}{\updefault}2}}}}
\put(5356,-7136){\makebox(0,0)[b]{\smash{{\SetFigFont{12}{14.4}{\familydefault}{\mddefault}{\updefault}1.5}}}}
\put(3691,-7136){\makebox(0,0)[b]{\smash{{\SetFigFont{12}{14.4}{\familydefault}{\mddefault}{\updefault}1}}}}
\put(2026,-7136){\makebox(0,0)[b]{\smash{{\SetFigFont{12}{14.4}{\familydefault}{\mddefault}{\updefault}0.5}}}}
\put(1876,-6886){\makebox(0,0)[rb]{\smash{{\SetFigFont{12}{14.4}{\familydefault}{\mddefault}{\updefault}0}}}}
\put(1876,-5666){\makebox(0,0)[rb]{\smash{{\SetFigFont{12}{14.4}{\familydefault}{\mddefault}{\updefault}0.2}}}}
\put(1876,-4446){\makebox(0,0)[rb]{\smash{{\SetFigFont{12}{14.4}{\familydefault}{\mddefault}{\updefault}0.4}}}}
\put(1876,-3227){\makebox(0,0)[rb]{\smash{{\SetFigFont{12}{14.4}{\familydefault}{\mddefault}{\updefault}0.6}}}}
\put(1876,-2007){\makebox(0,0)[rb]{\smash{{\SetFigFont{12}{14.4}{\familydefault}{\mddefault}{\updefault}0.8}}}}
\put(1227,-3687){\rotatebox{90.0}{\makebox(0,0)[b]{\smash{{\SetFigFont{14}{16.8}{\familydefault}{\mddefault}{\updefault}Concentration parameter $c$}}}}}
\end{picture}%